\documentclass[%
12pt,
showpacs,
 aip,
 amsmath,amssymb,
]{revtex4}

\usepackage{graphicx}
\usepackage{dcolumn}
\usepackage{bm}

\newcommand{\R}{\mbox{R}}
\newcommand{\Ra}{\mbox{Ra}}
\newcommand{\Nu}{\mbox{Nu}}
\newcommand{\Pra}{\mbox{Pr}}
\newcommand{\T}{\mbox{T}}
\newtheorem{proposition}{Proposition}[section]
\newtheorem{lemma}{Lemma}[section]
\def\u{\mbox{\boldmath $u$}}
\def\i{\mbox{\boldmath$\hat{i}$}}
\def\j{\mbox{\boldmath$\hat{j}$}}
\def\k{\mbox{\boldmath$\hat{k}$}}

\begin{document}

\preprint{AIP/123-QED}

\title{Internal heating driven convection at infinite Prandtl number}

\author{Jared P. Whitehead}
\email{jaredwh@umich.edu}
\author{Charles R. Doering}
\email{doering@umich.edu}
\altaffiliation[Also at ]{Center for the Study of Complex Systems, Department of Physics, and Michigan Center for Theoretical Physics, University of Michigan, Ann Arbor, MI}
\affiliation{ 
Department of Mathematics, University of Michigan, Ann Arbor, MI 48109-1034 USA
}%


\date{\today}

\begin{abstract}
We derive an improved rigorous lower bound on the space and time averaged temperature $\langle T\rangle$ of an infinite Prandtl number Boussinesq fluid contained between isothermal no-slip boundaries thermally driven by uniform internal heating.
A novel approach is used wherein a singular stable stratification is introduced as a perturbation to a non-singular background profile, yielding the estimate $\langle T\rangle \geq 0.419 \left[\R \, \log \R\right]^{-1/4}$ where $\R$ is the heat Rayleigh number.
The analysis relies on a generalized Hardy-Rellich inequality that is proved in the appendix.
\end{abstract}

\maketitle

%

\section{\label{sec:intro}Introduction}
Rayleigh-B\'enard convection, the buoyancy-driven flow of a fluid heated from below and cooled from above, is a fundamental paradigm of complex nonlinear dynamics, pattern formation, and turbulence.
An ongoing challenge for analysis, theory, computation, and experiment is to ascertain how the heat transport depends on the thermal forcing as gauged by a nondimensional Rayleigh number and the fluid's material properties (typically the Prandtl number, the ratio of kinematic viscosity to thermal diffusivity) \cite{AhGrLo2009}.
Rigorous bounds on heat transport in Rayleigh-B\'enard convection within the Boussinesq approximation were pioneered by Howard \cite{Ho1963} and subsequently elaborated by Busse \cite{Bu1969}.
Later, following the motivational work of Hopf \cite{Ho1941}, an alternative variational framework for bounds on turbulent transport of momentum, mass, and in the case of Rayleigh-B\'enard convection, heat, known as the `background method' was formulated \cite{DoCo1996}.
This is the approach we adopt here. 

In this paper we consider an infinite Prandtl number Boussinesq fluid contained between two rigid isothermal boundaries thermally driven by constant internal heating.
This model is inspired by convection in the Earth's mantle where  the Prandtl number is ${\cal O}(10^{24})$ and the motion is predominantly driven by a semi-uniform heating from radioactive decay.
For definiteness we consider an isoviscous fluid subject to no-slip isothermal vertical boundary conditions (without loss of generality $T=0$ on the boundaries) with uniform heating in the bulk, an idealization of the actual geophysical conditions.
At low heating rates as measured by the dimensionless `heat Rayleigh number' $\R$, the fluid remains at rest and heat is transported to the boundaries by conduction with a parabolic temperature profile across the layer.
At high $\R$ convection sets in, increasing the effective thermal conductivity of the layer and consequently lowering the bulk averaged temperature.
The enhancement of heat transport is indicated by $\langle T\rangle$, the ratio of the bulk averaged temperature to that of the parabolic profile resulting from conduction alone.
The challenge is to determine how  $\langle T\rangle$ varies with $\R$.
The no-convection conduction solution exists (be it stable or unstable) for all $\R$ and realizes the upper limit on the bulk averaged temperature, $\langle T\rangle \le 1$.
At high $\R$ the question is, how low can $\langle T\rangle$ go?

This problem was previously considerd by Lu {\it et al}~\cite{LuDoBu2004} who used estimates originally derived for boundary driven convection \cite{DoCo2001} and a simple piecewise linear background profile to produce a lower bound on the space and time averaged temperature relative to that of the conduction state.
That result was 
\begin{equation}\label{Lu_bound}
\langle T\rangle \geq 0.81 \, \R^{-2/7}.
\end{equation}
Because $2/7 = 0.2857 \dots$, this is not inconsistent with a scaling law measured from direct numerical simulations \cite{SoLa1999} suggesting that
\begin{equation}\label{SoLa_bound}
\langle T \rangle \sim 1.65 \, \R^{-0.234}.
\end{equation}
We note, however, that those computations employed `free-slip' boundary conditions on the velocity field rather than the no-slip conditions employed in the analysis.
Boundary conditions for traditional Rayleigh Benard convection can drastically affect the dynamics so the comparison must be taken with a  degree of caution.

More recent developments \cite{IeKePl2006, DoOtRe2006} indicate that background profiles including some stable stratification in the bulk may be optimal for infinite Prandtl number convection suggesting room for improvement for rigorous results.
In particular, a singular integral analysis was performed to obtain a key estimate that was then utilized in the background method to produce an upper bound on the Nusselt number $\Nu$, the dimensionless measure of the enhancement of heat transport in boundary-driven Rayleigh-B\'enard convection, in terms of the traditional Rayleigh number $\Ra$ of the form $\Nu \lesssim [\Ra \log(\Ra)]^{1/3}$ \cite{DoOtRe2006}.  
Here we show that that key estimate is in fact a modified Hardy-Rellich inequality and we derive the sharp prefactor.
The newly derived inequality is then applied to the internal heating problem via the background method and, along with some additional considerations, we prove
\begin{equation}\label{new_bound}
\langle T \rangle \geq 0.419 \, \R^{-1/4}  \left(\log \R\right)^{-1/4}.
\end{equation}

The rest of this paper is organized as follows.
The next section describes the Boussinesq equations of motion with internal heating and provides an outline of the background method applied to the problem.
Section III introduces the particular background temperature field as a logarithmic perturbation of a quadratic profile and applies the modified Hardy-Rellich estimate to obtain the bound \eqref{new_bound}.
Section IV discusses these results and briefly remarks on the parallels between the internal heating and boundary driven convection problems.
The new derivation of the Hardy-Rellich inequality is described in the appendix.

\section{Internal Heating and the Background Method}
In non-dimensional variables the Boussinesq approximation of the Navier-Stokes equations with a constant internal heat source is
\begin{eqnarray}\label{finitePr_mom_eqn}
\frac{1}{\Pra}\left(\frac{\partial\u}{\partial t} + \u\cdot \nabla\u\right) + \nabla p &=& \nabla^2\u + \R \, \T \, \k \\ \label{heat_eqn}
\frac{\partial \T}{\partial t} + \u \cdot\nabla \T &=& \nabla^2 \T+1\\ \label{incomp_eqn}
\nabla \cdot \u &=& 0
\end{eqnarray}
where $\u = \i u + \j v + \k w$ is the velocity field, $T$ is the temperature field \cite{LuDoBu2004}.  
The heat Rayleigh number is $\R$ and $\Pra$ is the Prandtl number.
We consider periodic horizontal boundary conditions in all variables with no-slip conditions ($\u = 0$) at the bottom ($z=-1$) and top ($z=0$) of the box.
Combined with the incompressibility condition \eqref{incomp_eqn}, this implies that $\partial w/ \partial z = 0$ at the top and bottom boundaries as well.
For definiteness we choose the special case where the boundary temperature is isothermal:
\begin{equation}\label{temp_BC}
\left.\T\right|_{z=-1} = 0 =  \left.\T\right|_{z=0}.
\end{equation}

Defining the space-time average of a function $f(x,y,z,t)$ as
\begin{equation}\label{average}
\langle f\rangle = \lim_{t\rightarrow\infty} \frac{1}{t} \int_0^t dt' \, \frac{1}{L_x}\int_0^{L_x} dx \, \frac{1}{L_y}\int_0^{L_y} dy \, \int_{-1}^0 dz \, f(x,y,z,t')
\end{equation}
(assuming that the limits exist) we are interested in obtaining a lower bound on the average temperature $\langle \T\rangle$ in terms of the Rayleigh heat number $\R$.
We focus on the infinite $\Pra$ limit of \eqref{finitePr_mom_eqn}, the rigorous validity of which has recently been established \cite{Wang2004}, so that the Navier-Stokes equations become the Stokes equations
\begin{equation}\label{infPr_mom_eqn}
\nabla p = \nabla^2\u + \R\T \k.
\end{equation}

To apply the background method, break the temperature field up into a background profile and fluctuations according to $\T(x,y,z,t) = \tau(z) + \theta(x,y,z,t)$ where $\tau(-1)=\tau(0)=0$ leaving the fluctuation $\theta(x,y,z,t)$ to satisfy homogeneous Dirichlet conditions at the top and bottom boundaries as well.
Applying this decomposition to the equations of motion and considering the space time average of both the momentum and properly weighted temperature equations we arrive at the following bound on the average temperature (see \cite{LuDoBu2004} for details of the derivation):
\begin{equation}\label{T_bound_eqn}
\langle \T\rangle \geq 2\langle \tau\rangle -\langle\left(\tau'\right)^2\rangle 
\end{equation}
as long as the quadratic (in $\theta$) functional
\begin{equation}\label{H_defn}
H = \langle|\nabla\theta|^2\rangle + \langle2\tau'w\theta\rangle
\end{equation}
is positive semidefinite among and temperature fluctuations and velocity fields satisfying the boundary conditions.
$H$ is quadratic in $\theta$ because, applying the curl operator twice to \eqref{infPr_mom_eqn}, we see that there is a linear albeit nonlocal instantaneous slaving of $w$ to $\theta$
 \begin{equation}\label{wtheta}
 \Delta^2 w = -\R\Delta_H\theta
 \end{equation}
 where $\Delta$ is the full Laplacian while $\Delta_H$ is the horizontal Laplacian in $x$ and $y$.
 
 For calculational convenience we apply the Fourier transform in the horizontal directions to obtain the relation for each wave number $k=|{\bf k}|$,
 \begin{equation*}
 \left(\frac{d^2}{dz^2}-k^2\right)^2 \hat{w}_{\bf k} = \R k^2 \hat{\theta}_{\bf k}
 \end{equation*}
 where now for all $\bf k$ the single-wavenumber quadratic forms
 \begin{equation*}
 H_{\bf k} := \int_{-1}^0 \left[\left|\frac{d\hat{\theta}_{\bf k}}{dz}\right|^2 + k^2|\hat{\theta}_{\bf k}|^2 + 2\tau'Re[\hat{\theta}_{\bf k}\hat{w}_{\bf k}] \right] dz
 \end{equation*}
 must all remain positive semidefinite.
 In the following we consider $H_{\bf k}$ wavenumber by wavenumber so we may drop the $\hat{\cdot}$ and subscript ${\bf k}$.
In other words we seek to maximize $2\langle \tau\rangle -\langle\left(\tau'\right)^2\rangle$ while maintaining positivite-semidefiniteness of
 \begin{equation}\label{H_defn_wavenumber}
H := \int_{-1}^0 \left[\left|\frac{d\theta}{dz}\right|^2 + k^2|\theta|^2 + 2\tau'Re[\theta w]\right]dz
\end{equation}
uniformly in $k$, where $\theta(z)$ satisfies Dirichlet boundary conditions and $w(z)$ solves
\begin{equation}\label{wtheta_wavenumber}
w'''' - 2k^2w'' + k^4w = \R k^2\theta,
\end{equation}
while satisfying both Dirichlet and Neumann boundary conditions on on $[-1,0]$.

Previous analysis of this problem and boundary driven convection considered background profiles $\tau(z)$ constant in the bulk of the layer so the only possible negative contribution to $H$ relied on the product of $w(z)$ and $\theta(z)$ in boundary layers where both are constrained to be relatively small in magnitude.
As was shown for boundary driven convection, however, a stably stratified (i.e. $\tau'(z) > 0$) profile in the bulk can be exploited to utilize the positive weighted correlation between $w$ and $\theta$ resulting from the slaving and improve the positivity of $H$, allowing for sharper estimates \cite{DoOtRe2006}.

\section{Singular perturbation of a stably stratified profile}
\begin{figure}
\includegraphics[width=19pc,angle=0,trim = 16mm 65mm 20mm 68mm, clip]{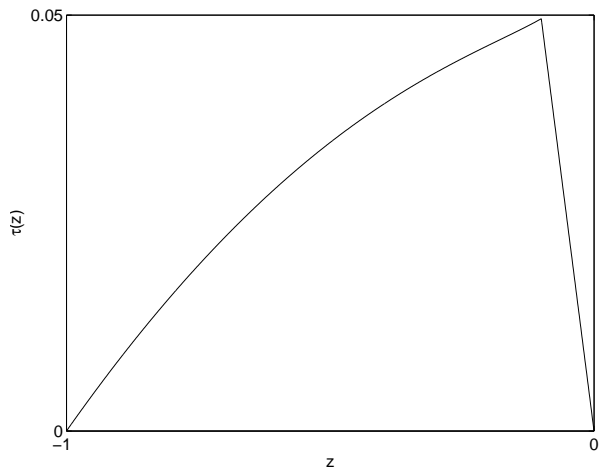}\caption{Background profile \eqref{final_profile}. }\label{fig:tau_final}
\end{figure}

Consider the family of background profiles illustrated in Figure \ref{fig:tau_final},
\begin{equation}\label{final_profile}
\tau(z) = \left\{\begin{array}{cc}
a\log\left(\frac{-1}{z}\right) + b\left(1-z^2\right) & -1\leq z\leq -\delta\\
\frac{-z}{\delta} \left[a\log(1/\delta) + b(1-\delta^2) \right] & -\delta \leq z\leq 0, \end{array}\right.
\end{equation}
where the positive parameters $\delta<1$, $a$, and $b$ will be chosen to optimize the bound.
The logarithmic term enhances the positivity of $H$, and hence leads to an improved scaling of the boundary layer with $R$, while the quadratic term is meant to increase the integral of $\tau(z)$ sufficiently to offset the slow logarithmic growth near $z=-1$ and lessening the negative impact of the Dirichlet integral in \eqref{T_bound_eqn}.

It is easily verified that
\begin{eqnarray} \label{integral1}
\int_{-1}^0\tau(z)dz &=& a\left[1-\delta-\frac{\delta\log(1/\delta)}{2}\right]+b\left[\frac{2}{3}-\frac{\delta}{2}-\frac{\delta^3}{6}\right]\\ \label{integral2}
\int_{-1}^0\left(\tau'(z)\right)^2dz &=& a^2\left(\frac{1}{\delta}-1+\frac{\left[\log(1/\delta)\right]^2}{\delta}\right)\\
&+& ab\left(2\frac{\log(1/\delta)}{\delta}+4-2\delta\log(1/\delta)-4\delta\right)\\
&+& b^2\left(\frac{1}{\delta}+\frac{4}{3}-2\delta-\frac{\delta^3}{3}\right)
\end{eqnarray}
producing the lower bound on the average temperature $\langle \T\rangle$ given by \eqref{T_bound_eqn} when the positivity of $H$ is maintained.
An appropriate choice of the scaling of the parameters $a$ and $b$ with respect to $\delta$ will allow us to determine both the `correct' boundary layer scaling, and to maximize the lower bound on $\langle \T\rangle$.

The two key inequalities required for the analysis are
\begin{eqnarray} \label{quad_estimate}
\int_{-1}^0\theta(z)w^*(z)z dz & \leq& 0\\ \label{new_estimate}
&\text{and}& \nonumber  \\
\int_{-1}^0\frac{\theta(z)w^*(z)}{z}dz &\leq& \frac{4}{\R}\int_{-1}^0\frac{|w(z)|^2}{z^3}dz \ \leq \ 0.
\end{eqnarray}
The first inequality (\ref{new_estimate}) is an exercise in integration by parts the details of which are left to the reader.
The second inequality \eqref{new_estimate} is a restatement---and slight improvement in the prefactor---of the key result previously derived for Rayleigh-B\'enard convection \cite{DoOtRe2006}.
While a prefactor improvement may be considered minor, our approach to prove it is simplified significantly and embeds the problem in the familiar context of a generalized Hardy-Rellich inequality.
The proof, provided in the Appendix, also indicates that the estimate with this prefactor is sharp.

To determine conditions guaranteeing the positivity of $H$ we reformulate it neglecting much of the $L^2$ norm of $\frac{d\theta}{dz}$ as well as the $k^2 | \theta |^2$ term and use \eqref{quad_estimate} to observe
\begin{eqnarray}
H &\geq& \int_{-\delta}^0\left|\frac{d\theta}{dz}\right|^2dz - 2a\int_{-1}^0\frac{Re[\theta w^*]}{z}dz\\
&-& \int_{-\delta}^0\left(\frac{2a\log(1/\delta)}{\delta}-\frac{2a}{z}-4bz+\frac{2b(1-\delta^2)}{\delta}\right)Re[\theta w^*]dz.
\end{eqnarray}
In the above we added the bulk terms to the boundary layer in order to apply \eqref{quad_estimate} and \eqref{new_estimate} to the entire interval.
Applying \eqref{new_estimate} then implies
\begin{equation}\label{H_star}
H \geq \int_{-\delta}^0\left|\frac{d\theta}{dz}\right|^2dz - \frac{8a}{\R}\int_{-1}^0\frac{|w|^2}{z^3}dz - \int_{-\delta}^0\left(\frac{2a\log(1/\delta)}{\delta}-\frac{2a}{z}-4bz+\frac{2b(1-\delta^2)}{\delta}\right)Re[\theta w^*]dz.
\end{equation}
Bound the magnitude of the last integral in \eqref{H_star} as follows:
\begin{eqnarray} \label{inner_integral}
& & \left|\int_{-\delta}^0\left(\frac{2a\log(1/\delta)}{\delta}-\frac{2a}{z}-4bz+\frac{2b(1-\delta^2)}{\delta}\right)Re[\theta w^*]dz\right|\\
&\leq& \int_{-\delta}^0 \left(\frac{2a\log(1/\delta)}{\delta}-\frac{2a}{z}-4bz+\frac{2b(1-\delta^2)}{\delta}\right)z^2\frac{|\theta|}{|z|^{1/2}}\frac{|w|}{|z|^{3/2}}dz\\ \label{sup_first_appears}
&\leq& 2\left(\sup_{-\delta\leq z\leq 0}\frac{|\theta(z)|}{|z|^{1/2}}\right)\left(\int_{-\delta}^0z^4\left[\frac{a\log(1/\delta)}{\delta}-\frac{a}{z}-2bz+\frac{b(1-\delta^2)}{\delta}\right]^2dz\right)^{1/2}\\
&\quad& \times \left(\int_{-1}^0\frac{|w|^2}{|z|^3}dz\right)^{1/2}.
\end{eqnarray}
The homogeneous boundary conditions on $\theta(z)$ mean that for $z\in \left(-\delta,0\right)$,
\begin{equation}
|\theta(z)| = \left|\int_z^0\frac{d\theta}{d\tilde{z}}d\tilde{z}\right| \leq |z|^{1/2}\left(\int_z^0\left|\frac{d\theta}{d\tilde{z}}\right|^2d\tilde{z}\right)^{1/2} \leq |z|^{1/2}\left(\int_{-\delta}^0\left|\frac{d\theta}{d z}\right|^2dz\right)^{1/2}.
\end{equation}
Hence we can bound the supremum in \eqref{sup_first_appears} and apply Young's inequality to see that
\begin{eqnarray} \label{final_inner_integral}
& & \left|\int_{-\delta}^0\left(\frac{2a\log(1/\delta)}{\delta}-\frac{2a}{z}-4bz+\frac{2b(1-\delta^2)}{\delta}\right)Re[\theta w^*]dz\right|\\\label{final_inner_integral}
&\leq& \int_{-\delta}^0\left|\frac{d\theta}{dz}\right|^2dz + \int_{-\delta}^0z^4\left[\frac{a\log(1/\delta)}{\delta} - \frac{a}{z}-2bz+\frac{b(1-\delta^2)}{\delta}\right]^2dz \times \int_{-1}^0\frac{|w|^2}{|z|^3}dz.
\end{eqnarray}
Inserting \eqref{final_inner_integral} into \eqref{H_star} we see that
\begin{equation}\label{final_H_bound}
H \geq \left\{\frac{8a}{\R}-\int_{-\delta}^0z^4\left[\frac{a\log(1/\delta)}{\delta}-\frac{a}{z}-2bz+\frac{b(1-\delta^2)}{\delta}\right]^2dz\right\}\int_{-1}^0\frac{|w|^2}{|z|^3}dz.
\end{equation}

The integral about the boundary layer in \eqref{final_H_bound} can be computed exactly.  At this point we choose $a=a'\delta/\log(1/\delta)$ and $b=b'\delta$ where $a'$ and $b'$ are ${\cal O}(1)$ absolute constants.
Then
\begin{eqnarray}
& & \int_{-\delta}^0z^4\left[\frac{a\log(1/\delta)}{\delta}-\frac{a}{z}-2bz+\frac{b(1-\delta^2)}{\delta}\right]^2dz\\
&=& a'^2\frac{\delta^5}{5}+2a'b'\frac{\delta^5}{5}+b'^2\frac{\delta^5}{5}+O\left(\delta^5\log(1/\delta)\right)
\end{eqnarray}
as $\delta\rightarrow 0$.
Comparing this with \eqref{final_H_bound} we see that the minimal requirement for $H$ to remain positive in the $\delta\rightarrow 0$ or $\R\rightarrow\infty$ limit is
\begin{eqnarray}
& & \frac{8}{\R} \sim a'\frac{\delta^4\left[\log(1/\delta)\right]}{5}+2b'\frac{\delta^4\log(1/\delta)}{5}+\frac{b'^2}{a'}\frac{\delta^4\log(1/\delta)}{5}\\
&\Rightarrow& \frac{1}{\R} \sim \frac{\xi(a',b')}{4}\delta^4\log(1/\delta)
\end{eqnarray}
where
\begin{equation}
\xi(a',b') = \frac{(a'+b')^2}{10a'}.
\end{equation}
This yields the scaling of the boundary layer thickness as
\begin{equation}
\delta \sim \left[\xi(a',b')\R\log(\R)\right]^{-1/4}.
\end{equation}

The average temperature is bounded by two times \eqref{integral1} minus \eqref{integral2} implying that, asymptotically,
\begin{eqnarray}
\langle \T\rangle &\geq& \frac{4}{3}b'\delta - a'^2\delta-2a'b'\delta-b'^2\delta\\
&\sim& \left(\frac{4}{3}b'-a'^2-2a'b'-b'^2\right)\xi(a',b')^{-1/4}\left(\R\log(\R)\right)^{-1/4}.
\end{eqnarray}
To obtain the `best' prefactor, we maximize over $a'$ and $b'$ to achieve
\begin{equation}\label{final_bound}
\left\langle T\right\rangle \geq \frac{2^{3/4}5^{1/4}}{6} \left(\R \log(\R)\right)^{1/4} \sim 0.419 \left(\R\log(\R)\right)^{1/4}
\end{equation}
where the optimal prefactor is obtained for $a'=\frac{1}{16}$ and $b'=\frac{7}{16}$.

\section{Discussion and conclusions}
The background profile \eqref{final_profile} can be considered the sum of a singular logarithmic profile and a smooth conduction-like quadratic profile.
If the logarithmic term only is considered, i.e., $b=0$, then the profile would be
\begin{equation}\label{log_only_profile}
\tau_0(z) = \left\{\begin{array}{cc} a\log\left(\frac{-1}{z}\right) & -1\leq z\leq -\delta\\
-\frac{a\log(1/\delta)z}{\delta} & -\delta\leq z\leq 0.
\end{array}\right.
\end{equation}
This would be analogous to the approach taken in \cite{DoOtRe2006} for boundary driven convection.  However if the same steps are followed one sees that the optimal bound occurs for $a \sim \frac{\delta}{\left[\log(1/\delta)\right]^2}$ in which case the bound becomes
\begin{equation}\label{log_only_delta_bound}
\langle \T\rangle \geq \frac{\delta}{\left[\log(1/\delta)\right]^2}.
\end{equation}
This bound on the averaged temperature is weaker than that derived previously.
However, the same analysis performed previously to ensure the positivity of $H$, but with \eqref{log_only_profile} as the background profile yields the scaling $\delta \sim \R^{1/4}$.
\begin{figure}
\includegraphics[width=19pc,angle=0,trim = 16mm 65mm 20mm 68mm, clip]{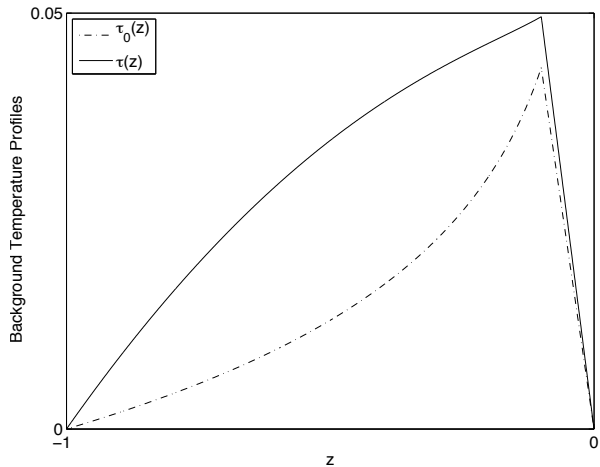}\caption{The background profile \eqref{final_profile} where $a$ and $b$ scale in the optimal sense compared to the logarithmic profile given by \eqref{log_only_profile}.}\label{fig:tau_compare}
\end{figure}

Fig. \ref{fig:tau_compare} yields some insight.
The purely logarithmic profile \eqref{log_only_profile}, depicted as the dashed plot, yields a thicker boundary layer at high $\R$ because the steep gradient near $z=0$ enhances the positivity of $H$ sufficiently to maintain the increased size of the boundary layer.  
But this costs dearly in the computation of the Dirichlet integral that negatively affects the estimate of $\langle\T\rangle$ while adding very little to the computation of $\langle \tau_0(z)\rangle$.
That is, while the quadratic term ($b>0$) thins the boundary layer, it also contributes significantly to $\langle\tau\rangle$ and raises the lower bound.

In previous applications of the background method \cite{DoOtRe2006, DoCo2001,OtWiWoDo2002} the scaling of the boundary layer dictates the bound: typically the heat transport is bounded by $\frac{1}{\delta}$ where $\delta$ is the size of the boundary layer.
Bounding the average temperature from below for the internal heating problem creates a different situation where the `optimal' boundary layer scaling in terms of $\delta$ yields a sub-optimal bound in terms of $\R$.

The bound \eqref{final_bound} is not inconsistent with numerical simulations \cite{SoLa1999} although we reiterate that the simulations employed stress-free (a.k.a. free-slip) boundary conditions on the velocity, as opposed to the no-slip conditions employed here, that may affect the scaling behavior.
The stress-free internal heating problem is addressed in \cite{WhDo2011b}.
It is also of interest to consider numerical solutions of the Euler-Lagrange equations for this problem, as has been done for the boundary driven convection for finite \cite{PlKe2003} and infinite \cite{IeKePl2006} Prandtl number Rayleigh-B\'enard convection.
Their solution would indicate what the true optimal background profile is, and may provide additional insight into the pursuit of further rigorous bounds.

\begin{acknowledgments}
\noindent
We thank Jeffrey Rauch for stimulationg discussions.  This work was supported in part by NSF Award PHY-0855335.
\end{acknowledgments}

\appendix

\section{A generalized Hardy-Rellich inequality}
We will establish \eqref{new_estimate} for all functions $w(z)$ and $\theta(z)$ that satisfy \eqref{wtheta_wavenumber} with the prescribed boundary conditions.  Note that with the change of variables $z\rightarrow -z$ this is equivalent to casting the problem on the positive unit interval as
\begin{equation}\label{positive_new_estimate}
Re \int_0^1\frac{\theta w^*}{z}dz \geq \frac{4}{\R}\int_0^1\frac{|w|^2}{z^3}dz
\end{equation}
where \eqref{wtheta_wavenumber} is satisfied for $z\in [0,1]$ and $w(0)=w(1)=w'(0)=w'(1)=\theta(0)=\theta(1)=0$.
In this context, \eqref{positive_new_estimate} is recognized as a factor of two improvement on the original proof \cite{DoOtRe2006}. 
As in the original proof we will prove the following proposition:

\begin{proposition}
If $0 < c \leq d \leq \infty$, the smooth function $w(z)$ satisfies
\begin{equation}\label{cd_BCs}
w(c)=0=w(d),~w'(c)=0=w'(d),
\end{equation}
and $\theta(z)$ is defined by $w'''' - 2k^2w'' + k^4w = \R k^2\theta$, then
\begin{equation}\label{cd_estimate}
Re\int_c^d\frac{\theta w^*}{z}dz \geq \frac{4}{\R}\int_c^d\frac{|w|^2}{z^3}dz.
\end{equation}
\end{proposition}

In order to see the connection between \eqref{cd_estimate} and Hardy-Rellich inequalities, make the change of variables $w(z) = z^{1/2}\zeta(z)$.  It follows that $\zeta(z)$ also satisfies \eqref{cd_BCs}.  Inserting this change of variables into the fourth order term that results from the definition of $\theta(z)$, we see that
\begin{equation}\label{w4w}
\int_a^b\frac{w''''w^*}{z}dz = \int_a^b|\zeta''|^2dz -\frac{3}{2}\int_a^b\frac{|\zeta'|^2}{z^2}dz + \frac{45}{16}\int_a^b\frac{|\zeta|^2}{z^4}dz.
\end{equation}
A similar calculation leads to
\begin{equation}\label{w2w}
\int_a^b\frac{w''w^*}{z}dz = -\int_a^b|\zeta'|^2dz + \frac{1}{4}\int_a^b\frac{|\zeta|^2}{z^2}dz.
\end{equation}
Putting \eqref{w4w} and \eqref{w2w} together, we see that \eqref{cd_estimate} can be restated as
\begin{lemma}
For smooth functions $\zeta(z)$ satisfying the boundary conditions \eqref{cd_BCs},
\begin{eqnarray} \nonumber
\int_c^d\left(|\zeta''|^2-\frac{3}{2}\frac{|\zeta'|^2}{z^2}+\frac{45}{16}\frac{|\zeta|^2}{z^4}\right)dz &+& k^2\int_c^d\left(2|\zeta'|^2-\frac{1}{2}\frac{|\zeta|^2}{z^2} \right)dz  +  k^4\int_c^d|\zeta|^2dz \ge \\ \label{HR_estimate}
&\geq& \ 4k^2\int_c^d\frac{|\zeta|^2}{z^2}dz.
\end{eqnarray}
\end{lemma}
Traditionally a Hardy-Rellich inequality is  formulated in terms of the $L^p$ norms of the operator $D^q = \frac{d^q}{dz^q}$ where $q=1,2$ and possibly higher orders (see \cite{KuPe2003} for example).  \eqref{HR_estimate} is, with the appropriate integrations by parts, nothing else than the $L^2$ norm of the differential operator $D^2-k^2$ acting on $\zeta(z)$.  The inclusion of the wave number $k$ here causes us to refer to this inequality as a generalized Hardy-Rellich inequality.

To prove the Lemma, consider the following one-parameter family of integrals,
\begin{equation}\label{quadratic}
0 \le \int_c^dz^{2\nu}\left[\left( D^2-k^2\right)\frac{\zeta}{z^\nu}\right]^2dz,
\end{equation}
where $\zeta(z)$ satisfies the homogeneous boundary conditions.  Expanding \eqref{quadratic} and integrating by parts multiple times leads to the following identity:
\begin{eqnarray*}
& & \int_c^d|\zeta''|^2dz + 2\nu(\nu-2)\int_c^d\frac{|\zeta'|^2}{z^2}dz + 2k^2\int_c^d|\zeta'|^2dz + \nu(\nu+6+\nu^3-4\nu^2)\int_c^d\frac{|\zeta|^2}{z^4}dz\\
&+& k^4\int_c^d|\zeta|^2dz \geq 2\nu^2k^2\int_c^d\frac{|\zeta|^2}{z^2}dz.
\end{eqnarray*}
Letting $\nu=\frac{3}{2}$ produces
\begin{eqnarray*}
& & \int_c^d|\zeta''|^2dz - \frac{3}{2}\int_c^d\frac{|\zeta'|^2}{z^2}dz + \frac{45}{16}\int_c^d\frac{|\zeta|^2}{z^4}dz + 2k^2\int_c^d|\zeta'|^2dz\\
&+& k^4\int_c^d|\zeta|^2dz \geq \frac{9}{2}k^2\int_c^d\frac{|\zeta|^2}{z^2}dz,
\end{eqnarray*}
which is easily rearranged to establish the Lemma.

The methodology employed above is not restrictive to this particular situation, and lends itself immediately to extension to higher order operators, and possibly higher dimensions as well.
The free parameter $\nu$ can be adjusted as desired, indicating a significant utility to this method of producing Hardy-Rellich type inequalities.
Hardy-Rellich inequalities with remainder terms can also be computed by optimizing over the wave-number $k$ (for an example of other Hardy-Rellich type inequalities with remainder terms see the work of Evans and Lewis \cite{EvLe2007}).

The strictness of the inequality derived here can be verified by considering functions $\zeta(z)$ that saturate \eqref{quadratic}, that is those functions satisfying the boundary conditions together with
\begin{equation}\label{saturating_equation}
\left(D^2-k^2\right)\frac{\zeta(z)}{z^\nu} = 0.
\end{equation}
Solutions of \eqref{saturating_equation} are linear combinations of modified Bessel functions:
\begin{equation}\label{saturating_function}
\zeta(z) = z^{1/2+\nu}\left[C_1 K_q(kz) + C_2I_q(kz)\right]
\end{equation}
where $q=\sqrt{2\nu^2+2\nu+1/4}$.
Just as the original Hardy inequality \citep{Hardy1920} is not saturated for any nontrivial analytic functions, the functions \eqref{saturating_function} cannot satisfy all the boundary conditions simultaneously so there is no analytic solution to \eqref{saturating_equation} that saturates \eqref{quadratic}.  Howeverm regularizing \eqref{saturating_function} appropriately at the boundaries will produce a sequence of functions that satisfy the boundary conditions and, in the unregularized limit, satisfy \eqref{saturating_equation}.
Hence while \eqref{quadratic} is never saturated, there can be no improvement on the prefactor derived by this method, i.e., the approach outlined here is not only robust and amenable to adaptation, but also produces sharp estimates.

\newpage


\end{document}